\documentclass[10pt]{iopart}
\usepackage{iopams}
\usepackage[dvips]{graphicx}
\newcommand{\ped}[1]{\ensuremath{_{\rm #1}}}

\begin{document}

\title[Independent determination of the two gaps in MgB$\ped{2}$]{Independent
determination of the two gaps  by directional point-contact
spectroscopy in MgB$_2$ single crystals}

\author{R.S. Gonnelli \dag\footnote[4]{To whom correspondence should be addressed
(gonnelli@polito.it)}, D. Daghero~\dag, G.A. Ummarino~\dag, V.A.
Stepanov~\ddag, J. Jun~\S, S.M. Kazakov~\S\ and J. Karpinski~\S}

\address{\dag\ INFM -- Dipartimento di Fisica, Politecnico di Torino,
Corso Duca degli Abruzzi 24, 10129 Torino, Italy}

\address{\ddag\ P.N. Lebedev Physical Institute, Russian Academy of
Sciences, Leninsky Pr. 53, 119991 Moscow, Russia}

\address{\S\ Solid State Physics Laboratory, ETH, CH-8093
Z\"{u}rich, Switzerland}

\begin{abstract}
Directional point-contact spectroscopy measurements were performed
for the first time in state-of-the-art MgB$_2$ single crystals.
The selective suppression of the superconductivity in the $\pi$
band by means of a suitable magnetic field allowed separating the
partial contribution of each band to the total point-contact
conductance. By fitting the partial conductance curves
$\sigma\ped{\sigma}(V)$ and $\sigma\ped{\pi}(V)$, we got an
independent determination of the two gaps, $\Delta\ped{\sigma}$
and $\Delta\ped{\pi}$, with a strong reduction of the experimental
uncertainty. Their temperature dependence was found to agree well
with the predictions of the two-band models for MgB$_2$.
\end{abstract}
\section{Introduction}
Even though a complete description of the many features of
MgB$\ped{2}$ discovered so far has not been achieved yet,  there
is a growing consensus on the fact that a large part of them can
be properly explained by admitting that two band systems are
present in MgB$_2$ \cite{twobands}, with two different energy gaps
\cite{Liu,Brinkman}.

One of the most convincing experimental supports of the two-band
model is the observation of two gaps by tunneling \cite{tunnel}
and point-contact spectroscopy (PCS) \cite{Szabo}, even though
there is a certain spread in the gap values obtained by different
groups. According to the model, the conductance of SIN or SN
junctions can be expressed as the weighed sum of two independent
conductances relevant to the two bands. Theoretical values for the
weight functions have also been calculated as a function of the
angle $\varphi$ between the direction of current injection and the
boron planes \cite{Brinkman}.

As far as the temperature dependence of the two gaps is concerned,
an inter-band pair scattering \cite{Suhl} is supposed to raise the
critical temperature of the $\pi$ bands so that both gaps
($\Delta\ped{\pi}$ and $\Delta\ped{\sigma}$) close at the same
temperature $T\ped{c}=39$~K \cite{Brinkman} even though their
amplitude is strongly different. While $\Delta\ped{\sigma}$
approximately follows a BCS-like curve (with non-standard gap
ratio $2\Delta\ped{\sigma}/k\ped{B}T\ped{c}=4.18$), a marked
reduction of $\Delta\ped{\pi}$ with respect to a BCS-like
behaviour (with gap ratio
$2\Delta\ped{\pi}/k\ped{B}T\ped{c}=1.59$) is expected at
T$\gtrsim$20~K.

Testing these predictions with high accuracy has been so far
impossible due to the lack of single crystals large enough to be
used for direction-controlled point-contact and tunnel
spectroscopy. In this paper, we present the results of the first
point-contact measurements in large single crystals of
MgB$\ped{2}$. We injected current along the $ab$ plane or along
the $c$ axis, and applied a magnetic field either parallel or
perpendicular to the $ab$ planes. This allowed us to separate the
partial contributions of the $\sigma$ and $\pi$ bands to the total
conductance, and to fit them obtaining the temperature dependence
of each gap with great accuracy. We will show that all the results
of this procedure confirm very well the predictions of the
two-band model appeared in literature.

\section{Experimental details}
The investigated MgB$_2$ single crystals were produced at
\emph{ETH} (Z$\mathrm{\ddot{u}}$rich). A mixture of Mg and B was
put into a BN container in a cubic anvil device. The crystals were
grown under a pressure of 30-35 kbar; the temperature was
increased during one hour up to the maximum of
1700-1800$^{\circ}$C, kept for 1-3 hours and decreased during 1-2
hours. This technique can give MgB$_2$ plate-like crystals up to
$1.5\times0.9\times0.2$~mm$^3$ in size and 200~$\mu$g in weight
\cite{crystals}, but the crystals used for our PCS measurements
were slightly smaller ($0.6\times 0.6 \times 0.04 $~mm$^3$ at
most). To remove possible deteriorated layers, the crystal surface
was etched with 1\% HCl in dry ethanol for some minutes.

Point contacts were made by using as a counterelectrode either a
small piece of indium pressed on the surface of the sample, or a
drop of Ag conductive paint. The contacts were positioned so as to
have current injection along the $ab$ planes or along the $c$
axis\footnote[7]{When the potential barrier at the interface is
small as in our case, the current is injected in a cone whose
angle is not negligible. In the ideal case of no barrier, this
angle is $\pi/2$. The probability for electrons to be injected
along an angle $\varphi$ in the cone is proportional to
$\cos{\varphi}$ \cite{Tanaka} so that it is maximum along the
normal direction anyway.}. Using such a non-conventional technique
for making point contacts was due to the very poor thermal
stability and reproducibility of the contacts made, as usual, by
pressing metallic tips against the sample surface. Instead, our
contacts had a remarkable mechanical stability during thermal
cycling, and showed very reproducible conductance curves. Even
though the apparent area of the contacts (about $10\times10\,
\mu\mathrm{m}^2$) was much greater than that required to have
ballistic current flow \cite{ballistic}, it must be borne in mind
that the \emph{effective} electrical contact only occurs in a much
smaller region. The resistance of our point contacts (that fell in
most cases in the range $10\div 50\, \Omega$) suggests that they
can be assumed to be in the ballistic regime.

\section{Results and discussion}
\begin{figure}[t]
\begin{center}
\includegraphics[keepaspectratio, width=0.9\textwidth]{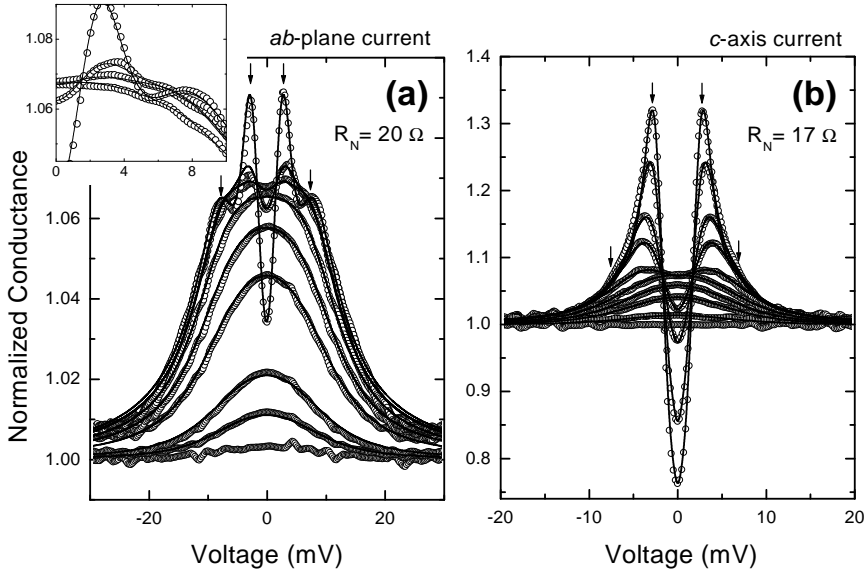}
\vspace{-5mm} \caption{(a) Some experimental conductance curves
(symbols) of a Ag-paste point contact with $ab$-plane current
injection at various temperatures from 4.2~K up to the critical
temperature of the junction, $T\ped{c}^{A}=34$~K. The curves are
normalized to the high-bias conductance \cite{nostroPRL}. Arrows
indicate the positions of the peaks in the low-temperature curves
($T=4.2$, 8.3, 11.9, 16.3~K), also shown in the inset. (b) Some
experimental conductance curves of a In-flake point-contact with
$c$-axis current injection. Arrows indicate the position of the
peaks and of the residual shoulder at higher voltage. Solid lines
in (a) and (b) are the BTK best-fitting curves.}
\end{center}
\vspace{-3mm}
\end{figure}

Figure 1 shows the temperature dependence of the experimental
\emph{normalized} conductance curves (circles) of point contacts
obtained by using an Ag-paint spot (a) and an In flake (b). The
values of the normal-state resistance (independent of the
temperature) are indicated in each panel. In (a) the current was
mainly injected along the $ab$ planes, while in (b) the current
direction was mainly parallel to the $c$ axis. The low-temperature
curves in (a) clearly show two peaks at $V=\pm 2.8$~mV and
$V\simeq \pm 7.6$~mV, that merge in a broad maximum above 15~K.
The low-temperature curves in (b), instead, show only a very sharp
peak at $V=\pm 2.8$~mV and a smooth shoulder at $V\simeq \pm
7$~mV. Clearly, these features are connected to the two gaps
$\Delta\ped{\pi}$ and $\Delta\ped{\sigma}$, and the different
relative amplitude of the peaks reflects the angular dependence of
the weight of each band in the total conductance. In the panels of
Figure 1, solid lines are the best-fitting curves calculated by
using the BTK model \cite{BTK} generalized to the case of two
gaps. In other words, the experimental normalized conductance was
fitted to a function of the form: $\sigma =
w\ped{\pi}\sigma\ped{\pi}+(1 -w\ped{\pi})\sigma\ped{\sigma}$. The
fit is almost perfect especially at low voltage. This might not
surprise since there are 7 adjustable parameters:
$\Delta\ped{\sigma}$ and $\Delta\ped{\pi}$, the broadening
parameters $\Gamma\ped{\sigma}$ and $\Gamma\ped{\pi}$, the barrier
transparency coefficients $Z\ped{\sigma}$ and $Z\ped{\pi}$, and
the weight of the $\pi$ band in the total conductance,
$w\ped{\pi}$. Actually, since neither the current direction nor
the contact resistance depend on the temperature, both
$w\ped{\pi}$ and the barrier parameters $Z\ped{\sigma}$ and
$Z\ped{\pi}$ were kept equal to their low-$T$ values.

\begin{figure}
\begin{center}
\includegraphics[keepaspectratio, width=0.6\textwidth]{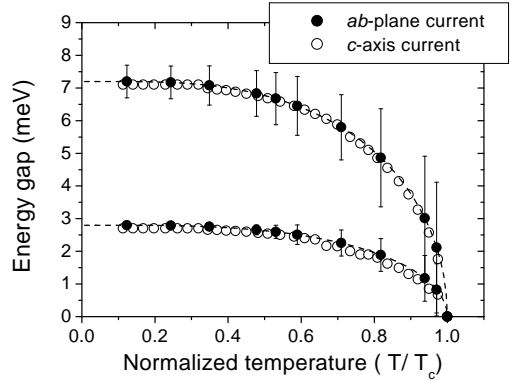}
\vspace{-2mm} \caption{Temperature dependence of the gap values
obtained from the fit of the curves in Figure 1a (solid circles)
and Figure 1b (open circles). Error bars are only shown for the
case of $ab$-plane current; in the case of $c$-axis current the
error on the small gap $\Delta\ped{\pi}$ is practically the same,
while the error on the large gap $\Delta\ped{\sigma}$ is even
larger than that shown, because of the smoothness of the related
feature in the conductance curves (see Figure 1b). Dotted lines
are BCS-like curves that best fit the experimental data.}
\end{center}
\vspace{-3mm}
\end{figure}

Figure 2 reports the temperature dependence of the two gaps given
by the fit, both for $ab$-plane current (solid circles) and for
$c$-axis current (open circles). The low-temperature gap values
$\Delta\ped{\sigma}=7.1\pm 0.5$~meV and $\Delta\ped{\pi}=2.9 \pm
0.3$~meV agree very well with the theoretical values
\cite{Brinkman} but are larger than those measured by specific
heat \cite{Bouquet} and thermal conductivity \cite{Sologubenko2}
in similar samples~\footnote{A possible reason is that the
theoretical gap values were calculated by starting from the
resistive $T\ped{c}$ (i.e., 39 K), while the bulk $T\ped{c}$ seems
to be lower \cite{campocritico2}. This might indicate that surface
effects play a role in MgB$_2$, affecting surface-sensitive
measurements such as point-contact spectroscopy.}. Unfortunately,
at higher temperature the error on the gap values increases so
much that it becomes practically impossible to determine whether
the $\Delta\ped{\pi}(T)$ and $\Delta\ped{\sigma}(T)$ curves follow
a BCS-like curve or not. Let us also mention that the values of
the parameter $w\ped{\pi}$ resulting from the fit are
$w\ped{\pi}=0.75 \pm 0.03$ in the case of $ab$-plane current and
$w\ped{\pi}=0.980 \pm 0.005$ in the case of $c$-axis current. The
values predicted by the two-band model are $w\ped{\pi}=0.66$ and
$w\ped{\pi}=0.99$ for current injection purely along the $ab$
plane and along the $c$ axis, respectively \cite{Brinkman}. Since
the crystal orientation is known with a few degrees of accuracy,
this apparent mismatch is actually due to the fact that, in
point-contact spectroscopy, the current injection occurs within a
finite solid angle \cite{Tanaka}.

A careful test of the two-band model obviously requires a more
accurate determination of the gap amplitude. This might be
obtained by reducing the number of free fitting parameters, e.g.
by separating the contributions of the two bands to the total
conductance. Some results obtained in polycrystalline samples
\cite{Szabo}, suggest that a magnetic field of suitable intensity
(namely, about 1~T at 4.2~K) can remove the gap in the $\pi$ bands
without affecting the gap in the $\sigma$ bands. We verified that
the same happens in our single crystals: we applied to each sample
(at $T=4.2$~K) magnetic fields of increasing intensity, either
parallel to the $c$ axis or to the $ab$ planes, and we observed
that in both cases the small-gap features in the conductance
curves vanish indeed when $B\simeq 1$~T \cite{nostroPRL}. The
effect of the field on the large gap depends instead on the field
direction. When $\mathbf{B} \parallel ab$-plane, the large-gap
features remain clearly distinguishable up to 9~T, with only some
marks of gap closing. Incidentally, this demonstrates that
$H\ped{c2}^{ab}>9$~T. When $\mathbf{B}
\parallel c$-axis, instead, the conductance peaks due to the large gap
merge together at $B\geq 4$~T giving rise to a broad maximum.
Based on the high value of $H\ped{c2}^{ab}$ at low temperature
(suggested both by our measurements and by other results on
similar samples \cite{campocritico2,campocritico1}), and supported
by some measurements at various temperatures
\cite{nostroPhysicaC}, we assumed that a field of 1~T parallel to
the $ab$ planes was too weak to affect seriously the large gap,
even at temperatures rather close to $T\ped{c}$.
\begin{figure}[t]
\begin{center}
\vspace{-1.6cm}
\includegraphics[keepaspectratio, width=\textwidth]{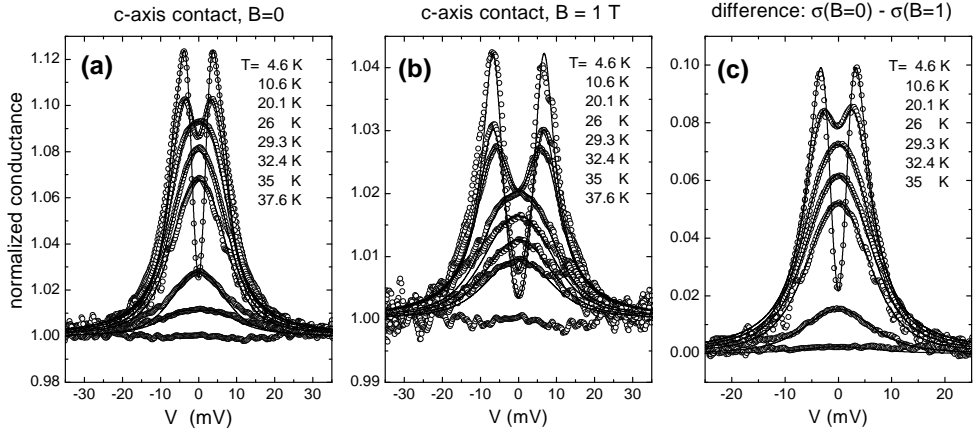}
\vspace{-2cm} \caption{(a) Some experimental normalized
conductance curves (open circles) measured in a In/MgB\ped{2}
point contact (R$\ped{N}\simeq 50\, \Omega$) with current
injection~along the $c$ axis, at different temperatures. Solid
lines are the best-fitting~curves obtained with the extended BTK
model. (b) Same as in (a) but with a magnetic field of 1~T
parallel to the $ab$ plane. Solid lines are the BTK best-fitting
curves, with three parameters: $\Delta\ped{\sigma}$,
$Z\ped{\sigma}$ and $\Gamma\ped{\sigma}$
($Z\ped{\sigma}(4.6\,\mathrm{K})=0.6$,
$\Gamma\ped{\sigma}(4.6\,\mathrm{K})\!=\!1.7$~meV).(c) Difference
between the total conductance (reported in (a)) and the partial
$\sigma$-band conductance (reported in (b)). Solid lines are the
BTK best-fitting curves, with three parameters: $\Delta\ped{\pi}$,
$Z\ped{\pi}$ and $\Gamma\ped{\pi}$
($Z\ped{\pi}(4.6\,\mathrm{K})=0.6$,
$\Gamma\ped{\pi}(4.6\,\mathrm{K})=2.0$~meV).}
\end{center}
\vspace{-3mm}
\end{figure}
Then, we measured the conductance curves of a In-MgB$\ped{2}$
point-contact in the $c$-axis-current case as a function of the
temperature, with no field (see Figure 3a) and with a field of 1~T
parallel to the $ab$ planes (see Figure 3b). If in the absence of
magnetic field the total \emph{normalized} conductance reads
$\sigma(B=0)=w\ped{\pi}\sigma{\ped{\pi}}+
(1-w\ped{\pi})\sigma\ped{\sigma}$, when the magnetic field
destroys the gap in the $\pi$ band it becomes:
$\sigma(B=1)=w\ped{\pi}+ (1-w\ped{\pi})\sigma\ped{\sigma}$. Thus
we fitted the curves measured in the presence of the field with
this function, that only contains three free parameters:
$\Delta\ped{\sigma}$, $\Gamma\ped{\sigma}$ and $Z\ped{\sigma}$. We
took $w\ped{\pi}=0.98$, that is the value obtained from the fit of
the total conductance at low temperature. Incidentally, the good
quality of the fit (solid lines in Figure 3) shows that the value
we got for $w\ped{\pi}$ is a very good evaluation, and that the
field completely suppresses the small gap. The temperature
dependence of $\Delta\ped{\sigma}$ obtained from this fitting
procedure is reported in Figure 4 (open circles).
\begin{figure}[t]
\vspace{-5mm}
\begin{center}
\includegraphics[keepaspectratio, width=0.6\textwidth]{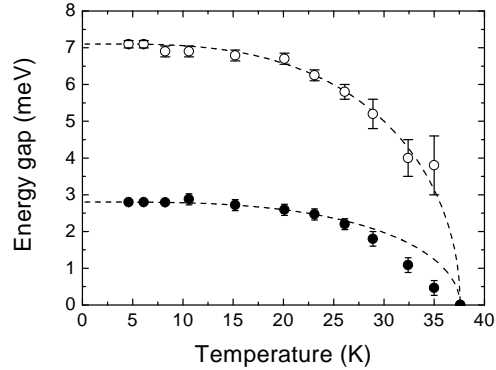}
\vspace{-2mm} \caption{Temperature dependence of the two gaps
obtained from the fit of the partial conductance of each band.
Open circles: $\Delta\ped{\sigma}(T)$ from the fit of the curves
in Figure 3b. Solid circles: $\Delta\ped{\pi}(T)$ from the fit of
the curves in Figure 3c. Dashed lines are the corresponding
BCS-like curves.}
\end{center}
\vspace{-3mm}
\end{figure}
A comparison with Figure 2 clearly shows the reduction of the
error bars and hence the improvement of the accuracy obtained by
using this novel technique. Also notice that
$\Delta\ped{\sigma}(T)$ follows rather well a BCS-like curve, as
in Figure~2. The absence of any field-induced suppression of
$\Delta\ped{\sigma}$ proves \emph{a posteriori} that a field of 1
Tesla parallel to the $ab$ plane has a negligible influence on the
$\sigma$ band, even at temperatures rather close to $T\ped{c}$.

An independent determination of the small gap can be obtained by
subtracting the conductance curve measured in the presence of the
field (Figure 3b) from those measured without field (Figure 3a).
The resulting curves are reported in Figure 3c. Notice that these
curves look particularly ``clean'' and noise-free since the
subtraction also allows eliminating some experimental fluctuations
that are present both in $\sigma(B=0)$ and in $\sigma(B=1)$. The
result of the subtraction can be expressed by the functional form
$\sigma(B=0)-\sigma(B=1)=w\ped{\pi}(\sigma\ped{\pi}-1)$. Fitting
the experimental data with this function allows determining the
three remaining free parameters $\Delta\ped{\pi}$,
$\Gamma\ped{\pi}$ and $Z\ped{\pi}$. The resulting temperature
dependence of the small gap is reported in Figure 4 (solid
circles). In this case, the error on the gap value is very small
even at high temperature, so that the deviation of the gap values
from the BCS-like curve (dashed line) results to be much larger
than the experimental uncertainty.

\section{Conclusions}
The selective removal of the gap in the $\pi$ bands from the total
conductance curves of point contacts is by itself a proof of the
existence of two distinct gaps in MgB$\ped{2}$. In this paper we
have shown that the use of single crystals allows a stricter test
of the two-band model. First, the fit of the total conductance
curves with the generalized BTK model gave us the weights of the
$\sigma$ and $\pi$ bands, which resulted in good agreement with
those predicted theoretically both for contacts along the $c$ axis
and along the $ab$ planes. Second, the separate analysis of the
partial conductances $\sigma\ped{\sigma}$ and $\sigma\ped{\pi}$
gave the highest-precision values of the gaps in MgB$\ped{2}$ ever
obtained by point-contact spectroscopy: at low $T$,
$\Delta\ped{\sigma}=7.1 \pm 0.1$~meV and $\Delta\ped{\pi}=2.80 \pm
0.05$~meV. While $\Delta\ped{\sigma}$ follows a BCS-like
temperature evolution, $\Delta\ped{\pi}$ deviates from the BCS
behaviour at $T>25$~K, in very good agreement with the two-band
model. Due to the small error on the gap value, this deviation is
here unquestionably determined for the first time.

This work was supported by the INFM Project PRA-UMBRA and by the
INTAS project ``Charge transport in metal-diboride thin films and
heterostructures''.

\end{document}